\documentclass[11pt]{article}
\usepackage{amsmath}
\usepackage{graphicx}
\usepackage{amssymb}
\usepackage{graphics,latexsym,amsfonts}
\usepackage{pstricks}
\usepackage{epigraph}
\usepackage{picture}
\usepackage[english]{babel}
\usepackage{enumitem}
\usepackage{cite}
\usepackage{caption}
\usepackage{subfig}
\usepackage[font=small]{caption}

\begin{document}

\title{\bf Order Book, Financial Markets, \\ and Self-Organized Criticality}
\author{
Alessio Emanuele Biondo \footnote{
Department of Economics and Business, University of Catania, Italy.} ,
Alessandro Pluchino \footnote{
Department of Physics and Astronomy, University of Catania and INFN Sezione di  Catania, Italy.} , 
Andrea Rapisarda \footnote{
Department of Physics and Astronomy, University of Catania and INFN Sezione di  Catania, Italy.}}
\date{}
\maketitle

\noindent
\textbf{Abstract} 
We present a simple order book mechanism that regulates an artificial financial market with self-organized criticality dynamics and fat tails of returns distribution. The model shows the role played by individual imitation in determining trading decisions, while fruitfully replicates typical aggregate market behavior as the ``self-fulfilling prophecy". We also address the role of random traders as a possible decentralized solution to dampen market fluctuations.

\bigskip


\section{Introduction}

Financial markets are characterized by  the  interactions of many interconnected heterogeneous agents, who trade with each other and follow their own expectations with  feedback mechanisms. The resulting aggregate behavior shows complex features, unpredictability and the occurrence of extreme events. Socio-economic systems can be studied as complex entities, by means of methods and concepts coming from statistical and theoretical physics \cite{mantegna1999introduction, Helbing1,  Helbing}. Such an approach helps studying financial markets exploiting  the idea of behavioral heterogeneity, entailing a specific role for the interaction among market participants, in terms of imitation and individual psychology \cite{simon1957models, tversky1974judgment, kahneman1979prospect, barberis2003survey}. 

The behavioral variability of agents operating in a complex network structure endowed with different informative sets justifies the evidence that empirical phenomena of socio-economic systems need new and alternative approaches. As an example, the beneficial role of random strategies has been shown  in several recent papers for the efficiency of  socio-economic systems \cite{AAPeter1, AAPeter2, AAParl}, and in particular of financial markets \cite{noiJSP, noiPRE, noiPLOSONE, noiCP}. Agent-based models may play a key role in understanding complex economic dynamics, needed for innovative policy design\cite{gatti2011macroeconomics}.

The adoption of agent-based approaches in financial markets models, surveyed in refs.  \cite{lebaron2006agent} and \cite{lux2009economics}, has revealed to be very useful to study the complex interactions of \emph{different} individuals with \emph{different} behaviors, as for example in \cite{brock1997models, brock1997rational, brock1998heterogeneous, chiarella1992dynamics, chiarella2001asset, day1990bulls, franke1998cautious, hommes2001financial, lux1995herd, lux1998socio, lux1999scaling}. Very often, in this stream of studies, two types of investors are taken into account: fundamentalists and chartists. The former are traders with an eye on the fundamental value of assets; thus they decide whether to buy a share or not, by looking at its current price level and by comparing it with its fundamental values (that is, roughly speaking, almost always the present discounted value of future expected dividends). The latter are technical analysts, who decide their strategies by following trends and graphic dynamics of past prices on charts.  

In the existing literature, the imitative behavior of a trader has often been modeled by means of a switching oscillation  from fundamentalists to chartists or vice versa. Recently, a more realistic kind  of imitation has been proposed in \cite{noiPRE2015}, so that the imitation refers only to the trading decision, no matter which group the trader belongs to. Thus, the persuasive strength of information may induce, say, a chartist to imitate a fundamentalist without switching group. In the present paper, we go further along this direction  and more realistically propose that the imitation will regard the trading status, i.e. the decision either to buy or to sell or even to wait without trading at all. 

We will explicitly refer to herding phenomena deriving from information cascades between agents \cite{Bikhchandani-Hirshleifer-Welch} as the underlying mechanism of financial avalanches. Differently from other attempts to describe herding in financial markets \cite{alfaranoluxwagner, kononoviciusgontis13}, our approach considers the pressure coming from the accumulation of information, by recalling some features of a Self-Organized Criticality (SOC) model for  describing earthquakes dynamics\cite{olami}. Our present model builds up on \cite{noiPRE2015}, by adding an order book mechanism that determines the asset price by the matching of supply and demand, as in real  markets. In such a way, heterogeneous traders of different kinds interact by means of personal strategies decided according to information, imitation and prospective utility.  

There exists a very well established branch of literature dealing with the characteristics and the dynamics of order books. As a matter of fact, the order book design is the mechanisms that lets financial markets everyday life to develop: the way orders are placed and influence the current price, the way the bid-ask spread is canceled trade after trade, the size and the timing of the execution of orders, and so on. Examples of these studies are in \cite{Gopikrishnan-et-al2000a, Challet-Stinchcombe2001, Maslov-Mills2001, Bouchaud-etal2002, Potters-Bouchaud2003}. Different market mechanisms have been studied in the literature dealing with the market microstructure, such as in \cite{Garman1976, Kyle1985, Glosten1994, Biais-etal1997, O'Hara1997, Hasbrouck2007} among others. In the present paper we begin with a simple representation of the order book: limit order always executed at the best price, with the simplifying assumption that just one asset exists and that only 1-share orders can be placed. Moreover, we do not consider any cancelation of orders, because we  presume that at any time step every trader decides whether to buy or to sell and, consequentially, to submit the corresponding order.

In a previous study, \cite{noiPRE2015}, a self-organized criticality (SOC) model of financial markets has been presented with interacting agents and a contagion mechanism. There, the price formation was strongly based on an exogenous source of noise. The model here presented, instead, represents a first step towards a further advance, because it replaces the external noise with a more realistic microstructure of trading that induces the price formation. By combining together the influence of herding dynamics at the aggregate level and the orders matching, this model represents, as far of our knowledge, the first attempt to embed in a unique framework two fundamental aspects of real financial markets: aggregate contagion effects and individual orders placement. 

The paper is organized as  follows: in section 2 the model is described; in section 3 simulation results are discussed; in section 4, conclusions and some policy suggestions are presented.

\section{The OB-CFP Model}

This model builds on the CFP model presented in \cite{noiPRE2015}, and adds to it an order book mechanism to simulate the operation of a financial market. This is the reason why we named it \textit{Order-Book-driven Contagion-Financial-Pricing} (OB-CFP henceforth). In the next subsections, all the elements of the model will be described, from the definition of traders to their character, to the type of their interaction and so on, till the aggregate dynamics. At each time step, the model proceeds following this evolution: as a first step, all market participants form their individual (heterogeneous) expectations for the future price; in a second step, according to individual expected values, traders select their trading status (hold, buy or sell); as a third phase, all orders are organized in the order book which operates the matching for the transactions to be done; finally, the new aggregate asset price is reported as function of both the last trading price and the possible market imbalances (either excess demand or excess supply), as it happens in true markets. It will be shown in subsection 2.2, that the contagion mechanism comes into play when traders have to decide their status, in order to take into account the possibility that the infectivity of an euphoric or pessimistic perception of the market could generate information cascades of buying or selling behavior.    

\subsection{The Order Book Dynamics}

Let us consider an ideal financial market where only one asset exists and money has just an ancillary function for it to serve just for transactions regulation. The population consists of a given number $N$ of market participants, i.e. traders $A_i$ (with $i=1,...,N$). At the beginning of each simulation, they are endowed with an equally valued portfolio, composed by the same initial quantity of money $M_i=M$ ($\forall i$) and the same initial quantity of the asset $Q_i=Q$ ($\forall i$). The total wealth $W_i$ of each trader will be therefore defined as: $W_i=M_i+Q_i\cdot p_t$, where $p_t$ is the price of the asset at time $t$. Two groups of traders exist: \emph{(i)} fundamentalists, \emph{(ii)} chartists. However, a third category of traders, i.e. random traders, will be also considered to study  their influence in the market dynamics. At each time step, traders will behave differently, according to their group. The difference among them is not new: fundamentalists presume the existence of a  \emph{fundamental value} and believe that the market price dynamics will tend to it. Therefore, they form their expectations by considering the actual difference between the fundamental price $p_{f}$ (different for each trader and randomly chosen in the interval $[p_F-\theta, p_F+\theta]$, where $p_F$ is a fixed global fundamental price) and the last market price $p_{t}$: they will expect a rise (fall) in the market price whether $p_{f} > p_{t}$ ($p_{f} < p_{t}$). Of course, a stationary dynamics is expected in case of equality. Thus, they form their expected price for the asset according to
\begin{equation}
E[p^{f}_{t+1}]=p_t+\phi (p_f - p_t) + \epsilon
\end{equation}
The parameter $\phi$ is a sensitivity parameter that describes the expected speed of convergence to the fundamental price and $\epsilon$ is a stochastic noise term, randomly chosen in the interval $(-\sigma, \sigma)$, with $\sigma$ fixed at the beginning of simulations and extracted with uniform probability. In order to limit the number of parameters, we let the value of $\phi$ be fixed but, in principle, it can be different for each trader of this group.

A chartist decides her behavior according to her inspection of past prices. Therefore, the expected price of each trader belonging to this group is a function of past prices: in particular we adopt the average of last $T$ prices over a time window that is different for each chartist ($T\in[2,T_{max}]$). Thus, a chartist will form ``her" expected price as a function of the difference between the last market price and the average of past $T$-prices, $p_T$. More precisely,
\begin{equation}
E[p^{c}_{t+1}]=p_t+\frac{\kappa}{T} (p_t - p_T) + \epsilon
\end{equation}
Also in this case, we consider  the  sensitivity parameter  $\kappa$ as a constant, whereas  $\epsilon$ is a stochastic noise term defined as in Eq.(1).   

Finally, we consider also random-trading agents. Random traders are investors who do not care at all about either previous or fundamental values: these market participants decide randomly (with uniform probability) whether to buy, to sell or to hold, without forming any expectation for the value $p_{t+1}$.

After having calculated her own expected price, each trader decides her order type, i.e. either to buy, or to sell, or to hold, and assumes the corresponding status $S_i$ (bidder, asker or holder). In case the expected price is greater than the actual one, it is profitable to buy the asset because the expected value of the owned portfolio is correspondingly higher. On the contrary, if a trader has a bearish expectation she will sell. It is worth to notice that in the model a sensitivity threshold $\tau$ has been introduced in such a way that if the expected price is equal or sufficiently close to the last price, the trader will decide to hold on without setting any orders. Of course, traders who decide to buy must have a positive amount of money ($M_i>0$) and, similarly, those who decide to sell must have a positive amount of the asset ($Q_i>0$). 

Once the individual status has been decided, each trader sets her order in the book by choosing the preferred price for the transaction. Both in case of sales and purchases, the price chosen by each trader (personal bid price for bidders and personal ask price for askers) for the transcription in the order book is a function of the expectation that inspired the status of the same trader. More precisely, the personal bid price will be a real (positive) random number smaller than the minimum between the money amount of the bidder and her expected price, while the ask price will be a real (positive) random number between the expected asset price of the asker and the actual global asset price.       

 We keep the order mechanism as simple as possible and allow for  1-asset-share quantity orders only. Thus, each order for a $+1$ or $-1$ quantity is posted with its corresponding bid or ask price. Bid prices are ranked in decreasing order of willingness to pay: in such a way, the trader who has set the highest bid price (namely the \textit{best-bid}) will be the top of the list and will have the priority in transactions. Conversely, ask prices are ranked in increasing order of willingness to accept: the trader with the lowest willingness to accept (who sets the so-called \textit{best-ask}) will be the top of the list and will have the priority in transaction execution. Then, the matching is done by comparing the best ask and the best bid. The number of transactions $N_T$ that actually does occur  between askers ($N_a$) and bidders ($N_b$) strictly depends on such a comparison. Furthermore, we let a feedback mechanism exist so that, according to the existence of an unsatisfied side of the market (i.e. either bidders or askers who could not trade for missing counterparts), the price receives a proportional shift $\omega$. Thus, in case of an excess of demand (i.e. bidders are greater in number than askers and therefore some of them cannot trade the asset at the desired price) the asset price will be increased proportionally to the excess itself. Conversely, when askers are greater in number than bidders, the price is decreased proportionally to the excess of supply. 

\begin{figure}[t]
\centering 
{\includegraphics[width=0.9\textwidth]{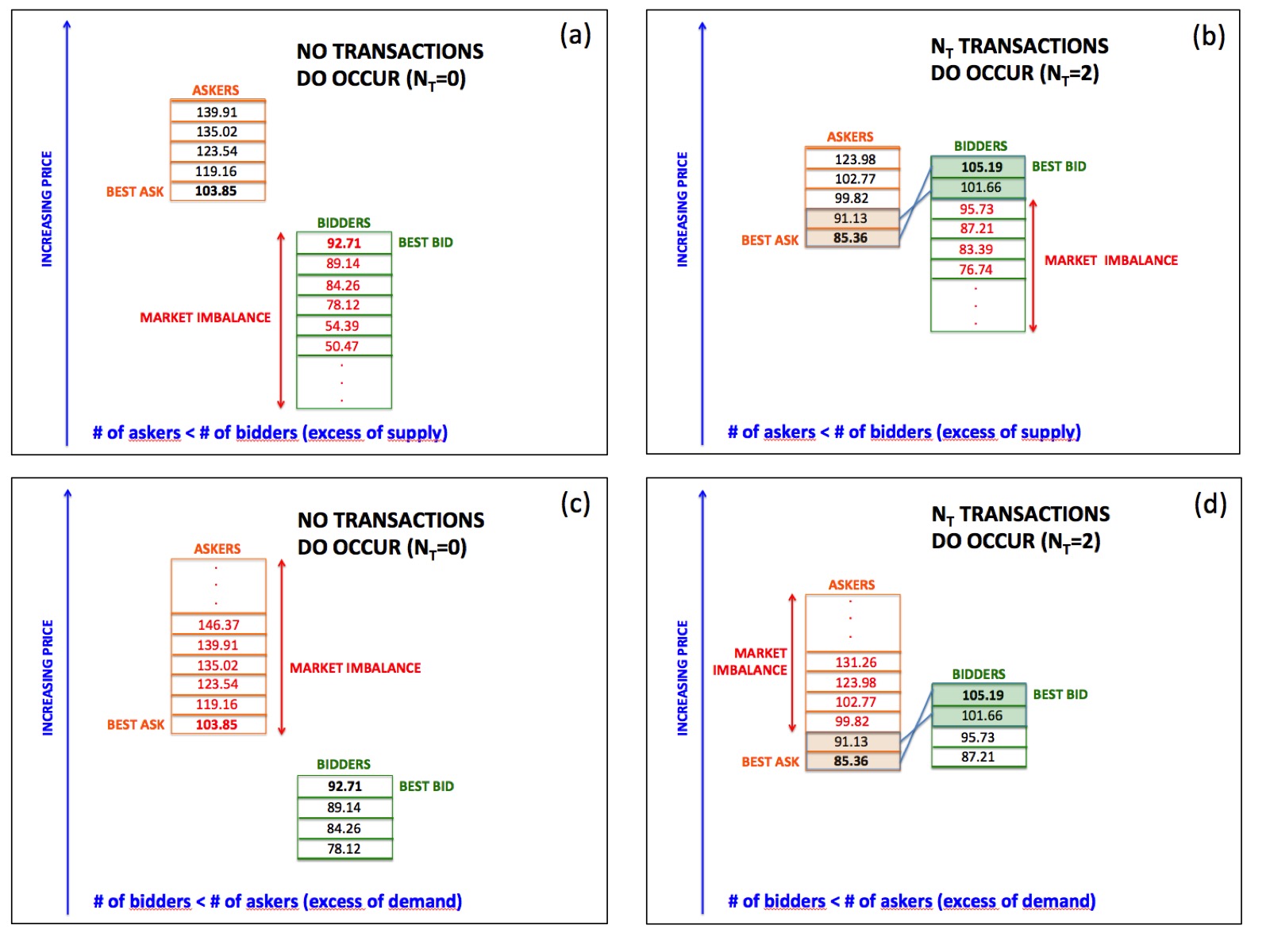}} 
\caption{\small{Examples of order book dynamics. No transactions are executed in cases depicted in panels \textit{(a)} and \textit{(c)} for missing counterparts. Transactions are executed only in cases depicted in panels \textit{(b)} and \textit{(d)}. In all cases, market imbalances will influence the future dynamics of the market price. See text for further details.}}
\end{figure}

Depending on all these variables at time $t$, several different cases must be considered in order to set the aggregate asset price at time $t+1$:

1) $N_a=0$ and $N_b>0$: no transactions occur and the new global asset price will be $p_{t+1} = p_t + \delta \cdot \omega$, where $\omega = N_b$ is the market imbalance and $\delta$ is a parameter which weights its effects on the new price;

2) $N_a>0$ and $N_b=0$: no transactions occur and the new global asset price will be $p_{t+1} = p_t - \delta \cdot \omega$, with $\omega = N_a$; 

3) $0<N_a<N_b$ but best-bid $<$ best-ask: no transactions occur, as shown in Figure 1(a), and the new global asset price will be $p_{t+1} = p_t + \delta \cdot \omega$, with $\omega = N_b$; 

4) $0<N_a<N_b$ and best-bid $>$ best-ask: a given number $N_T$ of transaction do occur, depending on the matching among ask and bid prices present in the order book, as shown in Figure 1(b); the first transaction occurs among traders who posted their own order at the best price, both from the demand or the supply side, then transactions continue following the order in the book (ascending for the ask list and descending for the bid list) until the bid price is greater than the ask price and all the transactions are regulated at the ask price; finally, the new global asset price will be $p_{t+1} = p_L + \delta \cdot \omega$, where $p_L$ is the ask price of the last transaction occurred and $\omega = N_b - N_T$; 

5) $0<N_b<N_a$ but best-bid $<$ best-ask: no transactions occur, as shown in Figure 1(c), and the new global asset price will be $p_{t+1} = p_t - \delta \cdot \omega$, with $\omega = N_a$; 

6) $0<N_b<N_a$ and best-bid $>$ best-ask: a given number $N_T$ of transaction do occur, depending on the matching among ask and bid prices present in the order book and following the same procedure described for the case $N_a<N_b$, as shown in Figure 1(d); finally, the new global asset price will be $p_{t+1} = p_L - \delta \cdot \omega$, where $p_L$ is the ask price of the last transaction occurred and $\omega = N_a - N_T$.

Notice that the new global asset price $p_{t+1}$ depends only on the previous price $p_{t}$ and on the market imbalance emerging from the order book dynamics. This is a more realistic description of the price formation with respect to eq. 4 of ref. \cite{noiPRE2015}, where an exogenous source of noise $\omega = \epsilon + e^{\beta I_{av}(t)}$ was fundamental in determining global price fluctuations. 

The next subsection is devoted to show how the price formation process is also affected by the endogenous herding dynamics among traders that exhibits self-organized criticality.

\subsection{The Aggregate Market}

At the aggregate level, in our artificial financial market traders $A_i$ (with $i=1,...,N$) are connected among themselves in a \emph{Small World} (SW) network \cite{noiPRE}, which is usually adopted in order to describe realistic communities in social or economical contexts. In particular, as shown in Figure 2, we consider here a $2$-dimensional regular square lattice with open boundary conditions and an average degree $<k>=4$. See ref. \cite{noiPRE} for more details. 

\begin{figure}[t]
\begin{center}
{\includegraphics[width=0.5\textwidth]{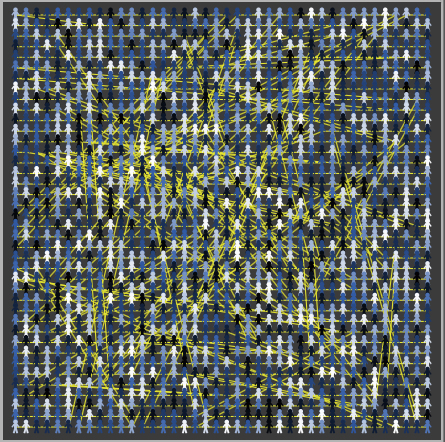}} 
\caption{\small{The 2D Small World network adopted in this model, with $N=1600$ agents (same figure as in ref. \cite{noiPRE2015}). Traders are distributed on a square lattice where short- and long-range links exist. Different colors represent levels of information: the brighter a trader is, the more informed she is. Initial levels of information are distributed randomly.}}
\end{center}
\end{figure}

Each agent in our simulated market receives two streams of informative pressures: a {\it global} one ($a$) and an {\it individual} one ($b$), \cite{noiPRE, noiCP, noiPRE2015}. 
\begin{itemize}
\item[($a$)] All traders receive a global informative pressure uniformly at every time-step from external sources. To each trader is associated a real variable $I_i(t)$ $(i=1,2,...,N)$ that represents her informative endowment at time $t$. Initially, at $t=0$, the informative level of each trader is set to a random value in the interval $[0,I_{th}]$, where $I_{th}=1.0$ is a threshold assumed to be the same for all agents. Then, at any time-step $t > 0$, the information accumulated by each trader is increased by a quantity $\delta I_i$, different for each agent and randomly extracted within the interval $[0,(I_{th}-I_{max}(t))]$; this may lead one or more traders to exceed their personal threshold value $I_{th}$, thus triggering the herding mechanism.  
\item[($b$)] On the other hand, every trader may receive, if involved in an avalanche,  a supplementary amount of information from her individual neighbors in the network, which is additive with regards to the global one ($a$); this may lead, again, a trader to accumulate enough information to exceed her personal threshold value $I_{th}$ and to trigger the herding mechanism. 
\end{itemize}

\begin{figure}[t]
\centering 
{\includegraphics[width=0.6\textwidth]{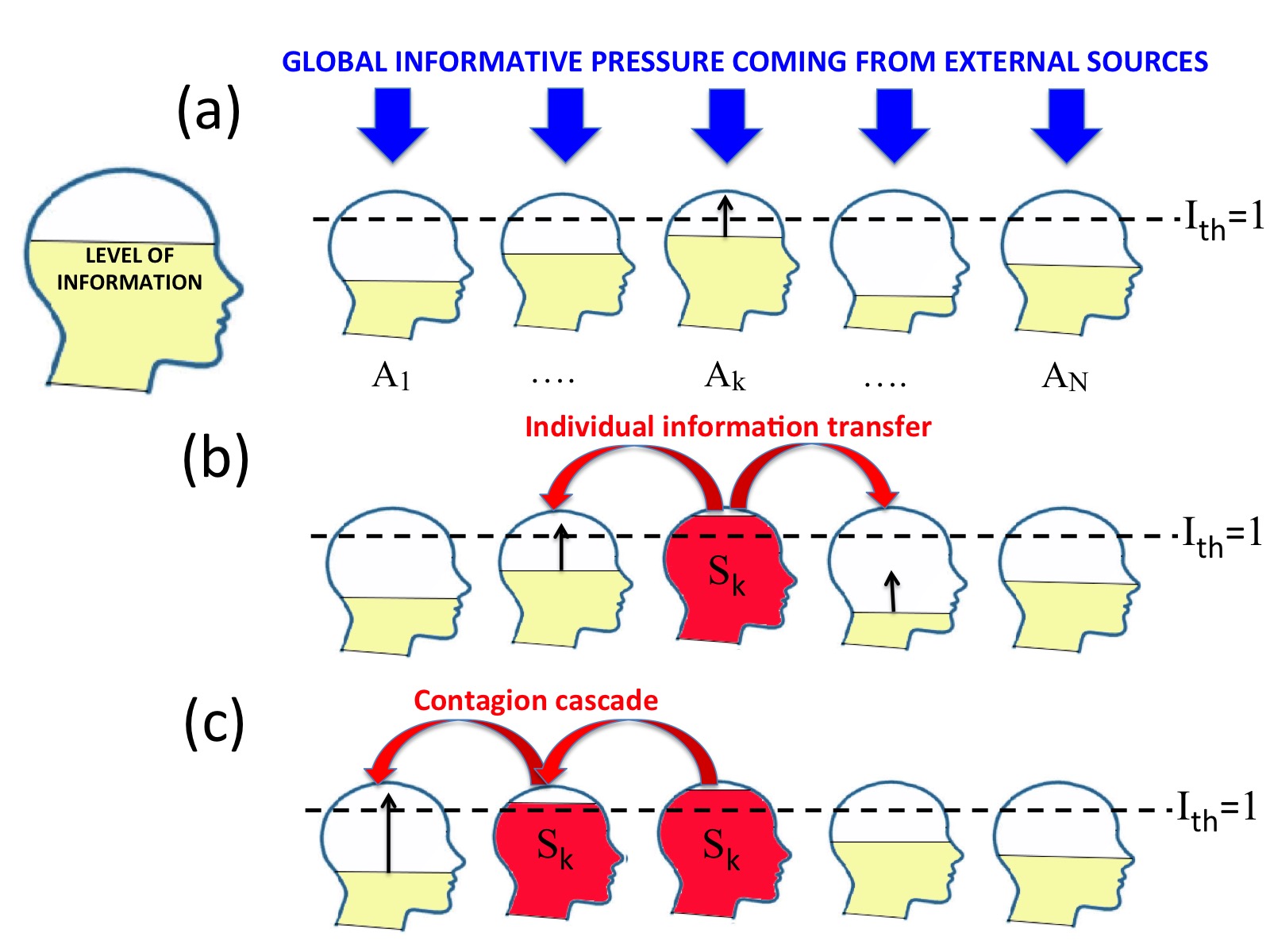}} 
\caption{\small{Informative pressure and contagion cascade.}}
\end{figure}

The herding mechanism is at the origin of the contagion effect and may involve exclusively the non-random traders.
On one hand, fundamentalists or chartists (i.e. non-random traders) are influenced by both the informative streams above described. When some non-random trader $A_k$ (either fundamentalist or chartist) surpasses her threshold at a given time $t=t_{av}$, see Figure 3(a), she becomes \textit{active} and transmits an informative signal about her status $S_k$ (asker, bidder or holder) to her neighbors within the trading network, see Figure 3(b). Such an information transfer happens according to the following simple herding mechanism \cite{noiPRE}, analogous to the energy transmission in earthquake dynamics \cite{olami}:

\begin{equation}   
\label{av_dyn}       
I_k > I_{th}  \Rightarrow \left\{ 
	\begin{array}{l}
       I_k \rightarrow 0, \\
       I_{nn} \rightarrow I_{nn} + \frac{\alpha}{N_{nn}} I_k,
       \end{array} 
	\right.
\end{equation}

where \textit{``nn''} denotes the set of nearest-neighbors of the active agent $A_k$. $N_{nn}$ is the number of direct neighbors, and the parameter $\alpha$ controls the level of dissipation of the information during the dynamics ($\alpha=1$ corresponds to the conservative case, but in our simulations we always adopted values strictly less than 1).  
As a consequence of the received amount of information, someone of the involved non-random neighbors may become active too and pass the threshold level as well: in this case, as shown in Figure 3(c), all the newly active traders will imitate the status $S_k$ of the first agent and will transmit, in turn, her own signal to their neighbors following again equation 3, and so on. In such a way, an informative avalanche will take place at time $t_{av}$, producing a contagion cascade of traders with the same status. It is worth to notice that, as it has been previously said, this kind of imitation does not imply a group switching of traders (from fundamentalists to chartists or vice versa), but only a change in their trading status. 

On the other hand, random traders are affected only by the external informative pressure. They neither influence other traders, nor are influenced by them. For this reason, as it has been shown in previous studies \cite{noiPRE, noiCP,noiPRE2015}, their role results to be crucial in damping the size of avalanches and reducing the contagion effect. 

In the next section such a herding dynamics and the order book mechanism will be combined, in order to adopt the complete OB-CFP model for several numerical simulations.

\section{Simulations results}

Each simulation has been computed by the following set of ordered steps.

Step 1. At $t=0$ we set the values for all the global parameters of the model: total number of traders ($N$), initial price ($p_0$), global fundamental price ($p_F$) and the corresponding variation ($\theta$), maximum extension for the chartists' time window ($T_{max}$), sensitivity of forecasts for fundamentalists ($\phi$) and chartists ($\kappa$), level of info dissipation ($\alpha$), weight of the market imbalance ($\delta$), intervals of variation for the stochastic noise terms ($\sigma$), sensitivity threshold for choosing the status ($\tau$); also the initial conditions for the individual parameters of the traders are set at this stage: information level, asset quantity, money and wealth. 

Step 2. The simulation starts with an opportune transient, during which the order book activity is suspended and the agents only receive global information from external sources and exchange individual information following their connections within the small-world lattice: this allows the system to reach, at a certain time $t_{SOC}$, the critical regime, where power-law distributed informative avalanches of any size do occur.        

Step 3. After the transient, i.e. for $t>t_{SOC}$, the order book dynamics starts to act, following the rules of section 2.1, and the global asset price time series emerges from the superposition of two simultaneous processes: the order book mechanism and the contagion due to SOC dynamics. The first one allows to determine the next asset price on the basis of both the matching between the two lists of bid and ask prices and the market imbalance due to the unsatisfied traders, while the second one affects the first by inducing herding cascades of (non-random) traders with the same status (asker, bidder or holder). The combination of these two processes is able to produce realistic fluctuations of the emerging aggregate price $p_{t}$, characterized by a non gaussian distribution of returns.

\begin{figure}[t]
\centering 
{\includegraphics[width=0.9\textwidth]{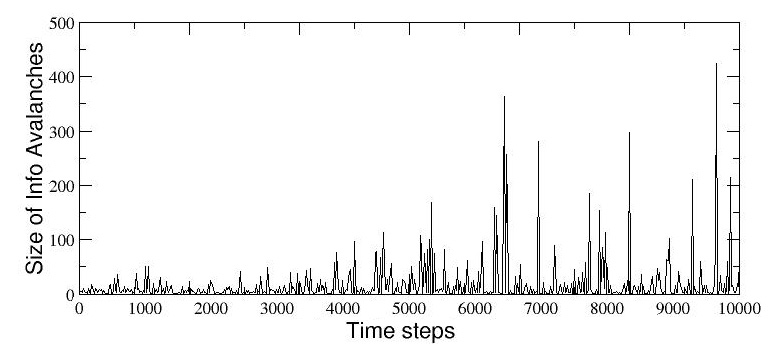}} 
\caption{\small{Time series of the informative avalanches size.}}
\end{figure}

First, we run a single-event simulation with a small-world lattice of $N=1600$ agents, divided in $800$ fundamentalists and $800$ chartists, and - at this stage - without random traders. We select the following, typical,  values for the global parameters: $p_0=100$, $p_F=120$, $\theta=30$, $T_{max}=15$, $\phi=2.0$, $\kappa=2.0$, $\alpha=0.95$, $\delta=0.05$, $\sigma=30$, $\tau=20$. We also set the initial conditions for the traders by randomly choosing the information level $I_i(0)\in[0,1]$ for $i=1,...,N$, and fixing the values of both the asset quantity $Q=50$ and the initial money amount $M=35000$ (in arbitrary units), equal for all traders. Thus,  agents' initial total wealth is $W_i=M+Q\cdot p_0=40000$, $\forall i$. 

From a single  run, limited to steps 1 and 2, one can  choose a suitable value for the transient duration $t_{SOC}$. As shown in Figure 4, $10000$ time steps are enough for the system to reach the critical-like state. Therefore we set $t_{SOC}=10000$ for all simulations. This transient time was not considered in the following numerical simulations and we took into account only the results coming from  simulations in the following critical regime   for further $20000$ time steps (steps 1-3).

\begin{figure}[t]
\centering 
{\includegraphics[width=0.95\textwidth]{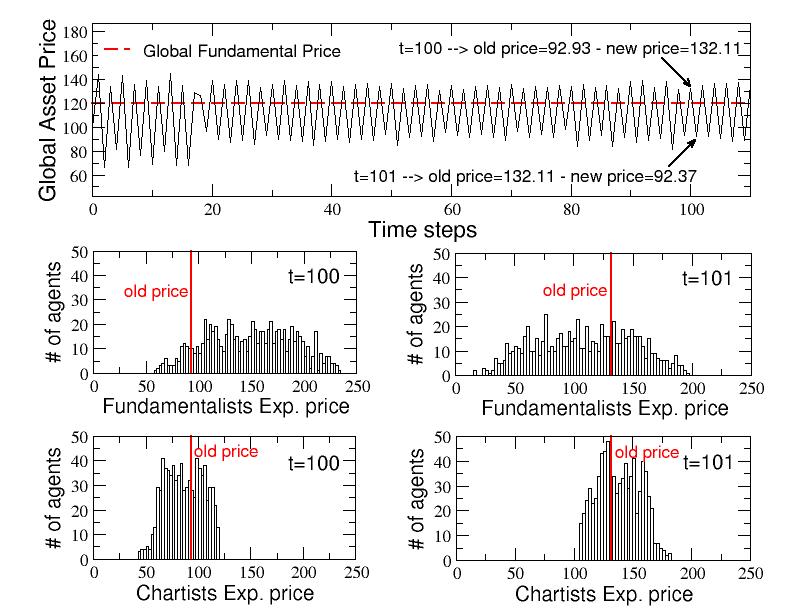}} 
\caption{\small{The first 110 values of the global asset price after the transient, i.e. for $t>t_{SOC}$. Smaller panels show the distribution of traders across expected prices, in both non-random categories.}}
\end{figure}

\subsection{Periodic price regime}

In the top panel of Figure 5 we show the evolution of the global asset price during the first 120 time steps after the transient (i.e. for $t>t_{SOC}$). It is clearly visible that, after around 20 time steps, the system rapidly settles down in a periodic regime, where the price oscillates around 112, an intermediate value between $p_0=100$ and $p_F=120$. 

Such a peculiar behavior can be intuitively explained by looking at the smaller panels of Figure 5, where the frequency distributions of the expectation prices for both fundamentalists and chartists are shown at two consecutive time steps, $t=100$ and $t=101$. Let us consider, first, the case $t=100$ (left column). Here, most of the expectation prices of fundamentalists (middle panel) are greater than the old global asset price $p_{99}=92.93$ (also reported in the panel as a vertical line), while the expectation prices of chartists (bottom panel) are more equally distributed around that value. This means that there will be much more bidders than askers ($N_b>>N_a$): therefore, the market imbalance (excess demand) will induce a substantial positive shift of the price, which will rise up to $p_{100}=132.11$. At the next time step, $t=101$ (right column), we find the opposite situation for the fundamentalists (middle panel), whose expectation prices are mostly below the global asset price $p_{100}$ (also reported in the panel as a vertical line), while the situation does not change for chartists (bottom panel), who are again equally distributed, in this case around the global price $p_{100}$. This will produce a substantial excess of supply ($N_a>>N_b$) that, in turn, will strongly push down the global asset price to $p_{101}=92.37$, bringing the system back to a condition of excess demand, and so on. One could also say that what happens here is a sort of (very realistic) self-fulfilling prophecy: the expectation of a price rise does indeed happen, as well as the expectation that it falls down. The result is the large periodic (and quite robust) oscillation of the price observed in the top panel. It is worth noting that, during this regime, the herding dynamics is not significant, since the large amplitude of the price oscillation hides the shifts due to the avalanche contagion effect.  

\begin{figure}[t]
\centering 
{\includegraphics[width=0.95\textwidth]{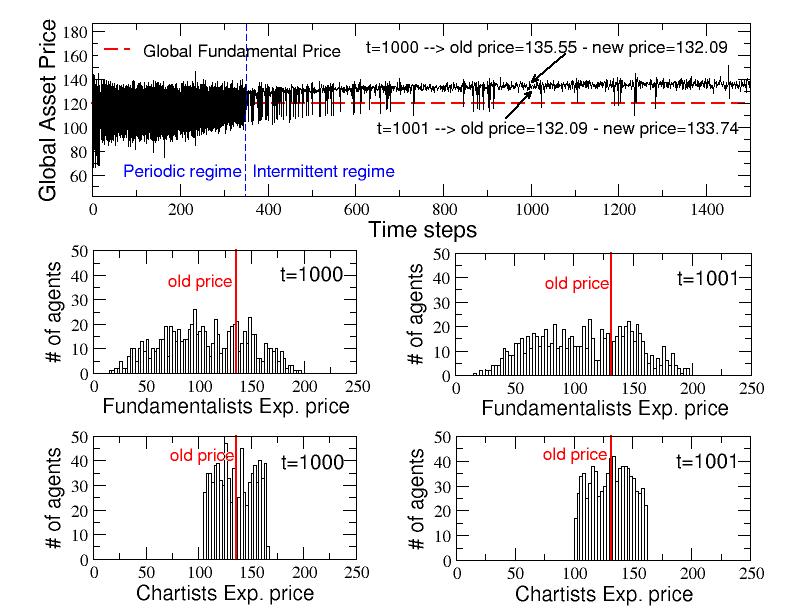}} 
\caption{\small{The first 1500 values of the global asset price after the transient, i.e. for $t>t_{SOC}$.}}
\end{figure}

\subsection{Intermittent price regime}
   
The periodic oscillation of the global asset price lasts for approximately 350 time steps. Then, suddenly, it gives way to a quite different behavior: as shown in the top panel of Figure 6, the amplitude of the oscillations undergoes a sharp reduction while its average value increases (from about 110 to about 135). Therefore, the herding dynamics can start to unveil its effects in the form of large and abrupt increments or decrements of the price, which give rise to a stable intermittent regime. 

As shown in the other (middle and bottom) panels of Figure 6, the strong reduction of the price oscillations for $t>t^*=350$ is essentially due to a sudden stabilization of the ranges of variation of the expectation prices for both fundamentalists and, in particular, chartists. Compared to the analogous distributions at $t=100$ and $t=101$ (shown in Figure 5), the distributions of the expectation prices at $t=1000$ and $t=1001$ are much more stable (always between 0 and 200 for fundamentalists and between 100 and 175 for chartists) and this, in absence of informative avalanches, dampens the market imbalance and, in turn, the price fluctuation. But when, sometimes, the contagion effect suddenly changes the status of a relevant number of traders, by inducing euphoria or panic in the network, the consequent excess of demand or supply gives rise to the price jumps that are visible in the intermittent regime.   

\begin{figure}[t]
\centering 
{\includegraphics[width=1.0\textwidth]{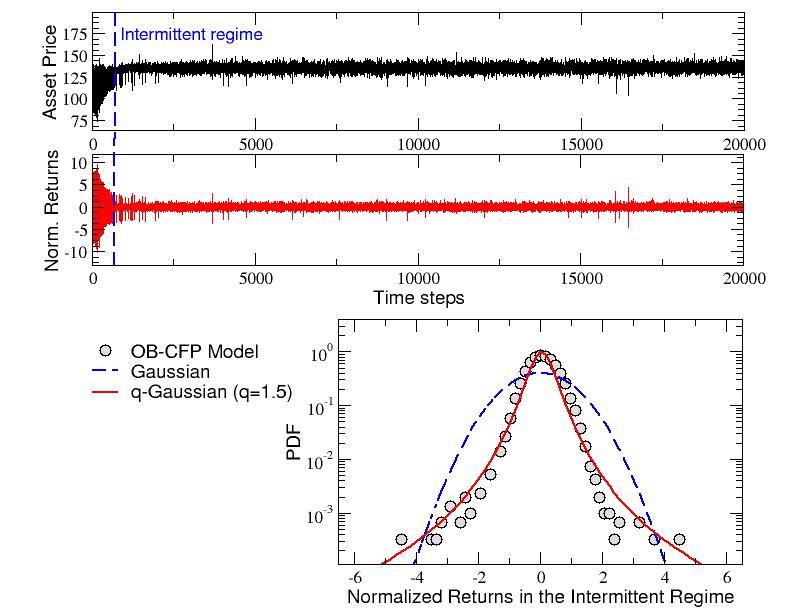}} 
\caption{\small{Complete time series of the global asset price after the transient (top panel), with the corresponding series of normalized returns (middle panel) and its probability distribution function (bottom panel) in the intermittent regime (circles). The dashed blue curve is a Gaussian with unit standard deviation, while the red curve which fits the data is a q-Gaussian with q=1.5. See text for further details.}}
\end{figure}

In the top panels of Figure 7 we finally plot the whole time series of the global asset price $p_t$, with $t=1,...,20000$, along with the corresponding series of returns $r_{t}=log(p_{t+1})-log(p_{t})$. In particular, we consider here normalized returns, defined as \mbox{$r^{NORM}_{t}=(r_{t} - r_{av})/r_{stdev}$}, where $ r_{av}$ and $ r_{stdev}$ are, respectively, mean and standard deviation calculated over the whole returns series. It is clearly visible that, as well as the price series, also the returns one is intermittent for $t>t^*$. This is confirmed by the plot of the probability density function (pdf) of normalized returns in the intermittent regime, shown in the bottom panel of Figure 7: in fact, compared with a Gaussian with unitary variance (dashed curve), the returns distribution (circles) shows a pronounced peak and the tendency to form fat tails, i.e. the typical shape of analogous distributions for real assets (see ref.\cite{noiPRE2015}). Data can be fitted by means of a $q$-Gaussian function (full line), defined as $G_{q}=A {[1-(1-q) B x^2]}^{1/(1-q)}$, with an entropic index $q=1.5$ ($A=0.98, B=7$), which measures the extent of the departure from Gaussian behavior (obtained in the limit $q=1$) \cite{tsallis}. Notice that this value of $q$ is very similar to the one obtained in a previous work \cite{noiPRE2015}, referred to  the returns distribution generated by the CFP model without the order book. Thus, this new model reveals to be able to exhibit the same stylized fact with a more realistic mechanism of price formation.

\begin{figure}[t]
\centering 
{\includegraphics[width=0.9\textwidth]{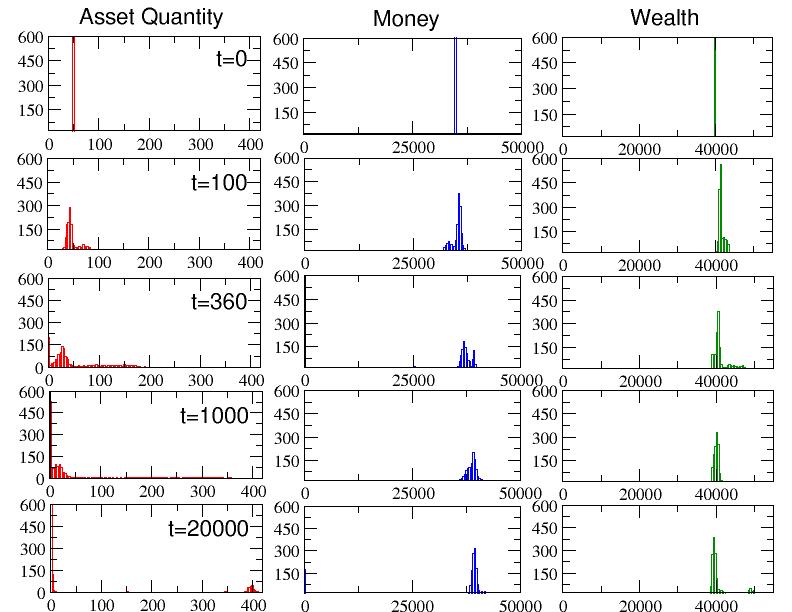}} 
\caption{\small{Distributions of the asset quantity (left column), of the money (central column) and of the total wealth (right column), of traders at different time steps (from top to bottom: $t=0$, $t=100$, $t=360$, $t=1000$ and $t=20000$).}}
\end{figure}

\subsection{Asset quantity, money and wealth distributions of traders}
     
Let us give a further look to the simulation run shown in Figure 7 but, now, from the point of view of the distributions of assets, money and - in turn - wealth, among the traders. Consider that, as already explained, agents without enough money $M_i$ cannot behave as bidders, while agents with zero quantity of the asset $Q_i$ cannot behave as askers. Therefore, in both cases, they cannot take part to the transactions and we expect that this should have visible consequences on the order book dynamics and on the global asset price evolution. In the left column of Figure 8, the distribution of the asset quantity is reported at different, increasing (from top to bottom), time-steps. Starting from the peaked initial histogram at $t=0$, when all the traders own the same asset quantity $Q=50$, such a distribution progressively spreads: the peak reduces and shifts on the left, while a tail arises and tends to stretch on the right. Around $t=t^*$ the peak touches the origin and an increasing number of agents start to lose all their assets ($Q_i=0$), while a small number of them accumulate assets until, at the end of the simulation, only about $200$ agents, all fundamentalists, remain with $Q_i>10$ (and only about $60$ of them with $Q_i>400$). Correspondingly, also the money distribution (middle column) spreads for $t>0$, and the initial peak centered at $M=35000$ starts to reduce and slightly shift to the right, while a thin tail of agents with decreasing money slowly diffuses to the left. Towards the end of the simulation (to be precise, at $t\sim15000$), the tail reaches the origin and an increasing number of traders start to lose all their money until, at the end of the simulation, one finds about $120$ agents with $M_i<100$ (in detail, they are all fundamentalists with $Q_i\sim 400$). Merging together asset quantity and money, the total wealth distribution (right column) shows a persisting peak of agents centered at the initial value $40000$, while just approximately one third of them  have, at the end, more wealth than they had at the beginning: in particular, only about $200$, all fundamentalists, end with more than $50000$. Summarizing, as one could expect, fundamentalists tend to be more conservative and to accumulate assets, even risking to lose money, but  maintaining anyway a good level of total wealth.     

\begin{figure}[t]
\centering 
{\includegraphics[width=0.9\textwidth]{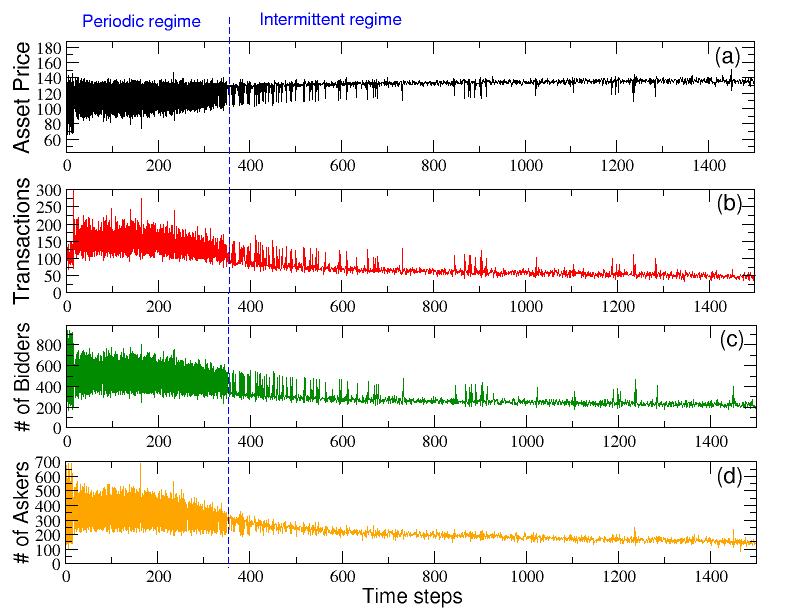}} 
\caption{\small{Comparison among the time evolutions of several quantities for 1500 time-steps: (from top to bottom) asset price (a), number of transactions (b), number of bidders (c) and number of askers (d). The transition between periodic and intermittent regimes, for our choice of the model parameters, does occur when the average number of transactions falls below 100, before settling around 50. }}
\end{figure}

It is interesting to notice that the rising of the peak of traders with $Q_i=0$ at $t\sim t^*$ (see the middle panel of the left column) is not a coincidence: actually, such a rapidly increasing number of agents without assets to sell causes a sudden reduction in the average number of transactions, which evidently destabilizes the periodic regime and induces, in turn, the onset of the intermittent one. As shown in Figure 9 (b), the average volume of transactions, which at $t=0$ is about $150$, will progressively decreases in time until, for $t > 350$, the number of transactions $N_T$ quickly settles in a plateau where it continues to oscillate around $50$ up to the end of the simulation. An analogous behavior can be observed for the number of bidders $N_b$ (c) and of askers $N_a$ (d), whose averages also decrease in time and tend to stabilize for $t>350$.  

\begin{figure}[t]
\centering 
{\includegraphics[width=1.0\textwidth]{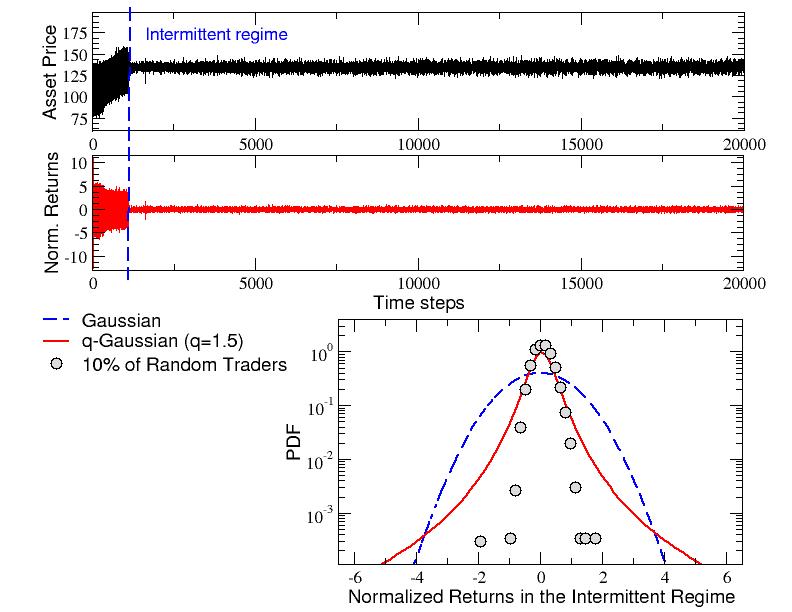}} 
\caption{\small{OB-CFP Model with 10\% of random traders. Time series of the global asset price after the transient (top panel), with the corresponding series of normalized returns (middle panel) and its probability distribution (bottom panel) in the intermittent regime. See text for more details.}}
\end{figure}

\subsection{The role of random traders}

In this last section, we study the effect of the introduction of a small percentage of random traders into the network, leaving unchanged all the other parameters, included the total number of agents ($N=1600$). In previous works \cite{noiPRE, noiCP,noiPRE2015} the role of random traders has been shown to be crucial in damping the herding avalanches and, therefore, in reducing the price fluctuations (and, thus, the volatility of the market): we confirm this finding  also in this case. In Figure 10 we show the results of a single run simulation, perfectly analogous to that one analyzed in the previous section (shown in Figure 7), but with the presence of $10\%$ of random traders in the artificial market. We have, thus, $160$ random traders, $720$ fundamentalists and $720$ chartists. The effect of the introduction of this new kind of agents, which decide at random their status (askers, bidders or holders) and do not take part to the herding process  is twofold: on one hand, as it is visible in both the top and middle panels of Figure 10, the presence of random traders delays the transition to the intermittent regime (which take place now at $t\sim 1200$); on the other hand, it also reduces the occurrence of large price fluctuations and, in turn, the shape of the returns distribution.  As it can be seen in the bottom panel of Figure 10, even if it is  still peaked and non Gaussian, such a distribution shows a strong reduction of the tails with respect to the analogous distribution obtained in absence of random traders (the q-Gaussian of Figure 7 is also reported for comparison). Thus, also in the context of the OB-CFP model, random-trading is confirmed to have a positive role in order to diminish  the price volatility in the market.

\section{Conclusions}
In this paper we have presented  a new model, namely the OB-CFP model,  based  an order-book-driven artificial financial market, with heterogeneous agents. The description of its realistic results has been provided, with particular regard to the fat tails of the returns distribution, the characterization of imitative behaviors and the ability to reproduce aggregate results that show compliance to the true feedback-driven market dynamics -a sort of self-fulfilling prophecy. In many ways this paper represents a fruitful extension of a previous study \cite{noiPRE2015} where a model of self-organized criticality was introduced in order to describe (and possibly control) the occurrence of crises and bubbles dynamics in financial markets. The numerical results presented confirm previous findings and, in particular, the beneficial role of random trading strategies. Here, we embedded an order book in such a way that the purchase/sell decisions may follow a more realistic allocative mechanisms. The basic innovation proposed in this paper is that, as for our knowledge, it is not common in literature to find a global market model that exhibits self-organized criticality behavior and embeds also the microstructure of trading by means of an operative order book. Although this first experiment relies on a simple mechanism, where trading of only one asset has been  considered,  further research will be devoted to remove this simplifying assumption in the orders settings, to obtain a more complete framework to question the applicability of efficient policies for market stabilization.

\section*{Acknowledgements}
This study was partially supported by the FIR Research Project 2014 N.ABDD94 of the University of Catania.

\end{document}